\documentclass[twocolumn,prl]{revtex4-1}

\usepackage{amsmath,amssymb,graphicx}

\usepackage{float}
\usepackage{amsthm}

\usepackage{color} 

\usepackage{bm}
\usepackage[symbol]{footmisc}
\usepackage{soul}

\usepackage{xcolor}

\usepackage{graphicx}

\newcommand{\mathsym}[1]{{}}
\newcommand{\unicode}[1]{{}}

\begin{document}
\def \beq{\begin{equation}}
\def \eeq{\end{equation}}
\def \bea{\begin{eqnarray}}
\def \eea{\end{eqnarray}}
\def \bes{\begin{split}}
\def \ees{\end{split}}
\def \besu{\begin{subequations}}
\def \esu{\end{subequations}}
\def \bea{\begin{align}}
\def \eal{\end{align}}
\def \bem{\begin{displaymath}}
\def \eem{\end{displaymath}}
\def \P{\Psi}
\def \Pd{|\Psi(\boldsymbol{r})|}
\def \Pds{|\Psi^{\ast}(\boldsymbol{r})|}
\def \Po{\overline{\Psi}}
\def \bs{\boldsymbol}
\def \dert{\frac{d}{dt}}
\def \k{\ket}
\def \br{\bra}
\def \bp{\hat b^+_{\Omega}}
\def \am{\hat a^-_{\omega}}
\def \ap{\hat a^+_{\omega}}
\def \pau {\partial_u}
\def \pav{\partial_v}
\def \paut{\partial_{\tilde u}}
\def \pavt{\partial_{\tilde v}}
\def \a{\alpha_{\Omega \omega}}
\def \b{\beta_{\Omega \omega}}

\title{Zel'dovich amplification in a superconducting circuit}
\author{Maria Chiara Braidotti$^{1,*}$, Andrea Vinante$^{2,3}$, Giulio Gasbarri$^{2}$, Daniele Faccio$^{1}$, Hendrik Ulbricht$^{2,}$} \email{mariachiara.braidotti@glasgow.ac.uk, h.ulbricht@soton.ac.uk}

\affiliation{$^{1}$School of Physics and Astronomy, University of Glasgow, G12 8QQ, Glasgow, UK.\\ 
$^{2}$Department of Physics and Astronomy, University of Southampton, SO17 1BJ, Southampton, UK.\\
$^{3}$Istituto di Fotonica e Nanotecnologie - CNR and Fondazione Bruno Kessler, I-38123 Povo, Trento, Italy.
}

\begin{abstract} 
Zel'dovich proposed that electromagnetic (EM) waves with angular momentum reflected from a rotating metallic, lossy cylinder will be amplified. However, we are still lacking a direct experimental EM-wave verification of this fifty-year old prediction due to the challenging conditions in which the phenomenon manifests itself: the mechanical rotation frequency of the cylinder must be comparable with the EM oscillation frequency. Here we propose an experimental approach that solves this issue and is predicted to lead to a measurable Zel'dovich amplication with existing superconducting circuit technology.
We design a superconducting circuit with low frequency EM modes that couple through free-space to a magnetically levitated and spinning micro-sphere placed at the center of the circuit. We theoretically estimate the circuit EM mode gain and show that rotation of the micro-sphere can lead to experimentally observable amplification, thus paving the way for the first EM-field experimental demonstration of Zel'dovich amplification.
\end{abstract}


\maketitle
In $1971$, Zel'dovich predicted the amplification of electromagnetic (EM) waves scattering off a spinning metallic cylinder~\cite{zel1971generation, zel1972amplification}, showing that the rotational energy of a spinning body can be transferred to the EM modes if the body spins rapidly enough, i.e. when
 \beq
\omega<q\Omega,
\label{eq_1}
\eeq 
where $\Omega$ is the cylinder rotation frequency and $\omega$ and $q$ are the frequency and the order of the angular momentum of the incident EM radiation.\\ 
The importance of this effect, aside from its own intrinsic interest, lies in the tight connection to other phenomena, from super-radiant scattering, i.e. amplification of waves from a rotating black hole as predicted by Penrose in 1969 \cite{penroseGravitational1969} to Hawking radiation i.e. evaporation of energy from a static black hole due to the interaction with quantum fluctuations \cite{hawking}.  Whereas laboratory analogues for these latter effects have been shown \cite{Rousseaux2008,Silke1,Silke2,jeff2,jeff3}, we are still lacking experimental verification of Zel'dovich's idea, as originally proposed with EM waves. This appears to be simply due to a technological difficulty in satisfying the condition in Eq.~(\ref{eq_1}), which nevertheless forces one to go back to carefully re-examine the underlying physical principles in order to propose a feasible experimental realisation. \\
Beyond Zel'dovich's original proposal, Bekenstein suggested to confine the EM mode in a cavity surrounding the metallic cylinder~\cite{bekenstein1998many} so as to resonantly increase the amplification effect. 
More recently Gooding et al.~\cite{gooding2019reinventing} proposed to impinge on the rotating disk from the direction of the rotation axis, therefore harnessing the  geometrical advantage of the dragging forces due to the penetration depth of the field inside the spinning body, as also suggested for acoustic experiments~\cite{faccio2019superradiant,Silke_sound}. 
However, this proposal also remains challenging~\cite{gooding2019reinventing}, requiring a macroscopic body to spin at tens of GHz.\\
  %
In this paper we describe a method to perform the original Zel'dovich experiment  aimed at observing amplification of EM waves from a conductive rotating body.\\
In more detail, we theoretically study the scattering from a rotating, levitated sphere of an EM mode that has angular momentum, i.e. in particular orbital angular momentum (OAM).
We show analytically how the amount of gain measured in the reflected EM mode depends on the  choice of the sphere material. We consider both a conductive non-magnetic particle and a non-conductive ferro-magnetic particle, showing that the predicted gain should be observable in a real experiment. \\
{\bf{The proposed system}} consists of a levitated conductive micro-sphere stably spinning at frequency $\Omega$ and a rotating EM mode used to probe the Zel'dovich effect.\\ 
Our proposed experimental set-up is schematically shown in Fig.~\ref{fig:1}. 
The EM field is the EM mode of a superconducting circuit that is
composed of 4 micro-coils of equal inductance $L$. These micro-coils are near-field coupled to the spinning, free-space (i.e. not in physical contact with the circuit) micro-sphere. The micro-coils are joined by transmission lines of equal length $\ell/4$ in a closed loop thus forming a closed circuit (see Fig. \ref{fig:1}). The chosen geometry, and in particular the length of wire between each coil, is such that adjacent micro-coils are phase-shifted by $\pi/2$ allowing propagating and counter-propagating normal modes of  EM frequency $\omega=c k_q$, where $c$ is the speed of light in vacuum, $k_q = 2 \pi q / \ell$, and non-zero OAM $q=1$ inside the circuit. \\
The whole system is placed in a cryogenic vacuum chamber at pressures of $\sim10^{-5}$ mbar to enable low temperature measurements at $T\sim4.2~$K (liquid helium temperature) or less, needed for superconducting technology and to obtain high electrical conductivities.\\ 
The initial driving to set the sphere in rotation is performed by switching the coil to an external circuit (not shown in Fig.~\ref{fig:1}). This driving magnetic field is generated by 4 waveform-generators, each synchronised with a $\pi/2$ phase shift, thus following the procedure demonstrated in~\cite{reichert2012complete,schuck2018ultrafast}. The driving field oscillates at a frequency $\omega_d$ and induces the micro-sphere to rotate at a frequency $\Omega\lesssim\omega_d$. $\Omega$ can also be tuned by changing the intensity of the driving field~\cite{reichert2012complete,schuck2018ultrafast}.
The maximum spinning frequency achievable with such a technique is only limited by the centrifugal forces overcoming the material stress limit and increases for decreasing sphere radius. For example, MHz rotation rates have been achieved by Schuck et al. with metallic spheres with a diameter of 0.5 mm~\cite{schuck2018ultrafast}.\\
Once the sphere reaches the desired rotation frequency 
the driving circuit and function generators are electronically switched off and probing is performed using the closed superconducting circuit shown in  Fig.~\ref{fig:1}. In a vacuum of $\sim10^{-5}$ mbar, we estimate that the levitated sphere will spin freely for more than one hour \cite{schuck2018ultrafast, gabis1996, epstein1924} (See Supplementary Material \cite{supply}), thus providing time to perform measurements with the superconducting circuit EM modes, under the assumption that these are significantly weaker than the driving EM field.\\
\begin{figure*}
\includegraphics[width=0.7\textwidth]{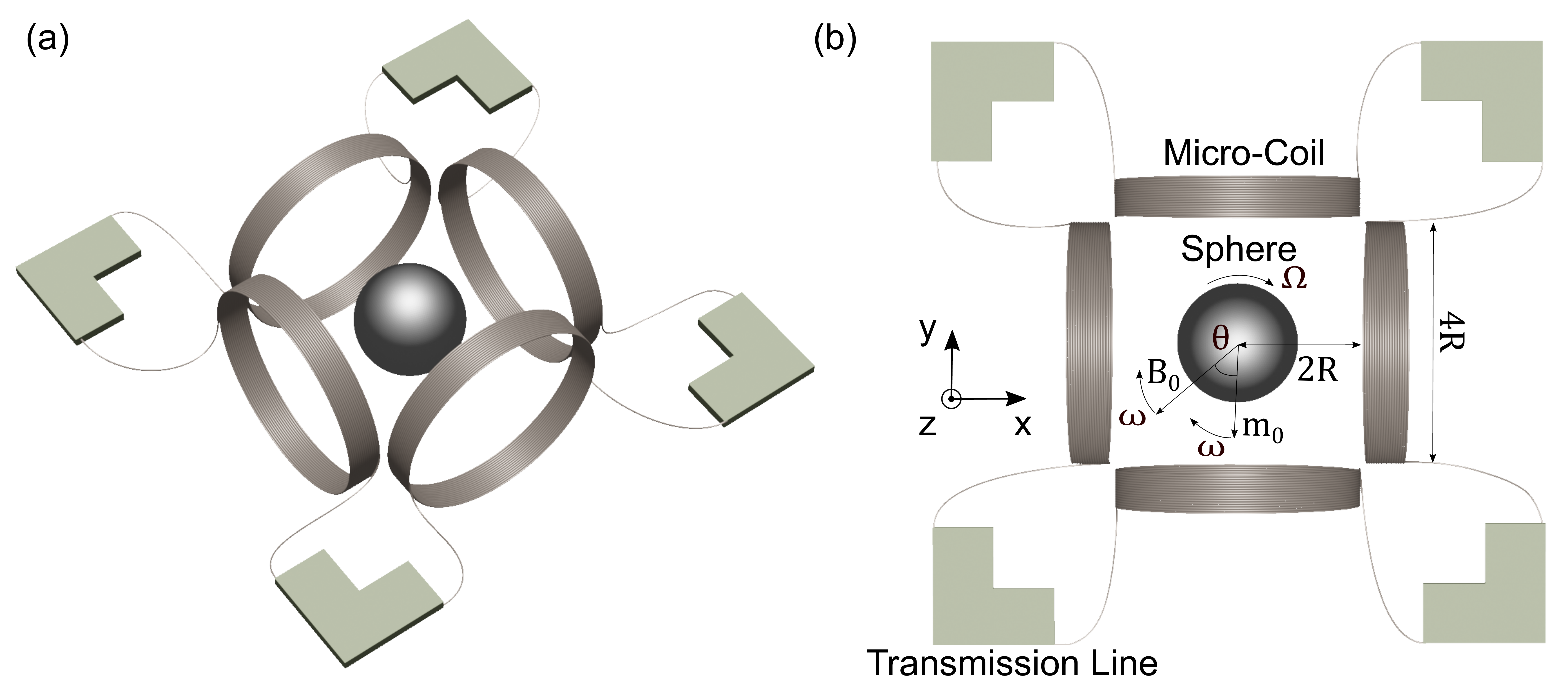}
\caption{(a) 3D Experimental set-up. (b) top view: the superconducting circuit is divided in $4$ sections of length $\ell/4$, connected by 4 inductors with inductance $L$ to form a closed ring configuration. The 4 coils have diameter equal to $4R$, where $R$ is the radius of the central sphere, rotating at frequency $\Omega$.  The distance between each coil and the center of the sphere is $2R$. The arrows denote the magnetic field $\bm{B_0}$ generated by the 4 coils and the momentum $\bm{m_0}$ induced on the sphere. Both vectors rotate at frequency $\omega$. $\theta$ is the angle between $\bm{B_0}$ and $\bm{m_0}$ and is related to the EM power dissipated in the sphere.}
\label{fig:1}
\end{figure*}
%
{\bf{Theoretical analysis of the proposed system.}} The probe EM field is generated by a sinusoidal current passing through the inductors. The current in the $j$-th lumped inductor, corresponding to a propagating mode in the superconducting circuit, can be expressed as: 
\begin{equation}
I_j \left( t \right) = \bar{I} e^{ i \left( k_q \xi_j - \omega t - \phi_0 \right)} ,
\end{equation}
where $\xi_j=j \ell/4$ denotes the coordinate of the position of each coil along the line 
and $j=(0,1,2,3)$. 
$\phi_0$ is an initial random phase and $\bar{I}$ is the peak current flowing in the coils due to the propagating wave.
The current flowing in the circuit can be measured by a SQUID, weakly coupled through a small inductance in series with the circuit (not shown in Fig.~\ref{fig:1}).\\
In the chosen configuration, the total probe magnetic field induced by the coils 
can be written by means of the Biot-Savart formula for a circular loop as \cite{Griffiths}:
\begin{equation}\label{B}
\bm{B_0} = 2 \beta \bar{I} \bm{b_0},  \quad \beta= \frac{N\mu_0}{8\sqrt{2}} {1\over R}    ,
\end{equation}
where $\bm{b_0} = \left(1,i,0\right)^Te^{i\omega t}$, $R$ is the radius of the micro-sphere, $N$ is the number of loops of each coil and the factor $2$ accounts for the contribution of the two opposite-facing coils. We observe that $\bm{B_0}$ is complex.  
In Eq.~(\ref{B}), we chose an arrangement of 4 micro-coils with radius $2R$ placed at a distance $2R$ from the center of the sphere (see Fig.~\ref{fig:1}b). This geometrical configuration has been chosen to maximise the coupling between the sphere and the coils. These parameters are also compatible with current technology: superconducting Niobium planar coils with $R\sim 100$ $\mu$m and high $N$, i.e. $N\sim 40$, can be fabricated using standard lithographic techniques and are typically used as input coils in conventional SQUIDs \cite{SquidBook}. \\
For simplicity the magnetic flux density $\bm{B_0}$ produced by the coils is assumed to be uniform over the microsphere volume, and the factor $\beta$ in Eq.~({\ref{B}}) is obtained by neglecting the thickness of each inductor. Thicker coils can also be considered, replacing the parameter $\beta$ by an effective $\beta_{\text{eff}}$ averaged over different loop. However, planar coils, having a large turn density, are the best candidates for this experiment \cite{planarcoils}. An example of planar coil that can be fabricated with existing technology showing $\beta_{\text{eff}}$ close to $\beta$ is reported in the SM \cite{supply}. 
%
We assume the spinning sphere axis is oriented along $z$, with the origin of the transverse $(x,y)$ plane set in the center of the sphere (see Fig.~\ref{fig:1}b).
In the reference frame co-rotating with the micro-sphere, the magnetic flux density can be written as $\bm{B_r}=2 \beta \bar{I} \bm{b_r}$ with $\bm{b_r} = \left(1,i,0\right)^Te^{i\omega_{r}t}$ and where $\omega_r=\omega-\Omega$ is the frequency of the field in the co-rotating reference frame. From the definition of $\omega_r$ we can see that  a negative co-rotating frequency will satisfy the Zel'dovich condition Eq.~(\ref{eq_1}) for $q=1$ that can be expressed as $\omega_r=\omega-\Omega < 0$: we will show that when $\omega_r < 0$, the effective dissipation becomes negative, implying a conversion of rotational mechanical energy into energy of the circuit electromagnetic mode.\\
The interaction between the magnetic probe field and the rotating sphere induces a magnetic dipole moment $\bm{m_0}$ on the sphere, which in the laboratory frame can be written as 
\beq
\bm{m_0}=\chi\bm{B_0}=(\chi'+i\chi'')\bm{B_0}  ,
\label{mom}
\eeq
where $\chi(\omega_r)$ is the complex response function of the sphere to the presence of the field in the co-rotating reference frame. The two vectors $\bm{m_0}$ and $\bm{B_0}$ rotate at the same EM frequency $\omega$ with an angular phase lag $\theta (\omega_r)$ determined by the imaginary part of the response function $\chi''(\omega_r)$ (see Fig.~\ref{fig:1}b), which gives rise to dissipation.
If the sphere is conductive and non-magnetic, $\chi$ depends on the electric conductivity $\sigma(\omega_r)$ whereas for a ferro-magnetic $(\sigma=0)$ sphere, $\chi$ is proportional to the complex relative permeability $\mu_r(\omega_r)$ 
(see details in the Supplementary Material \cite{supply}).\\
The EM power dissipated by the sphere can be calculated as:
\beq
W = \frac{1}{2} \Re\{ - \bm{m_0} \frac{d\bm{B_0^*}}{dt}\} = 4 \beta^2 \omega \chi''(\omega_r) I_0^2 ,
\label{power}
\eeq 
where the factor $1/2$ accounts for the average over one cycle, $\Re\{\cdot\}$ is the real part of the quantity in bracket, while $\bm{B_0}^*$ denotes the complex conjugate of the field~\cite{LandauBook}.
The susceptibility $\chi(\omega_r)$ is the Fourier transform of the real-valued linear response function $\chi(t)$, which implies that $\chi''(-\omega_r)= -\chi''(\omega_r)$~\cite{LandauBook_stat}. A direct and key consequence of this is that when the Zel'dovich condition Eq.~(\ref{eq_1}) is fulfilled, the power dissipated by the sphere becomes negative, i.e. power is radiated from the sphere into the EM circuit, leading to EM amplification. \\
The EM amplification can also be viewed as a result of the vector $\bm{m_0}$ preceding the vector $\bm{B_0}$, when $\omega_r<0$ (rather than following it as usually happens when $\omega_r>0$), as seen in the change of sign of rotation of the vector $\bm{m_r}$ in the co-rotating frame. Under these conditions the torque applied by the EM field, given by $\bm{T}=\Re\{\bm{m_0}\}\times\Re\{\bm{B_0}\}=-|\bm{B_0}| |\bm{m_0}| sin(\theta)\hat{z}$, also becomes negative so that the EM field is extracting mechanical energy from the spinning sphere. 

\begin{figure}[t]
\includegraphics[width=0.48\textwidth]{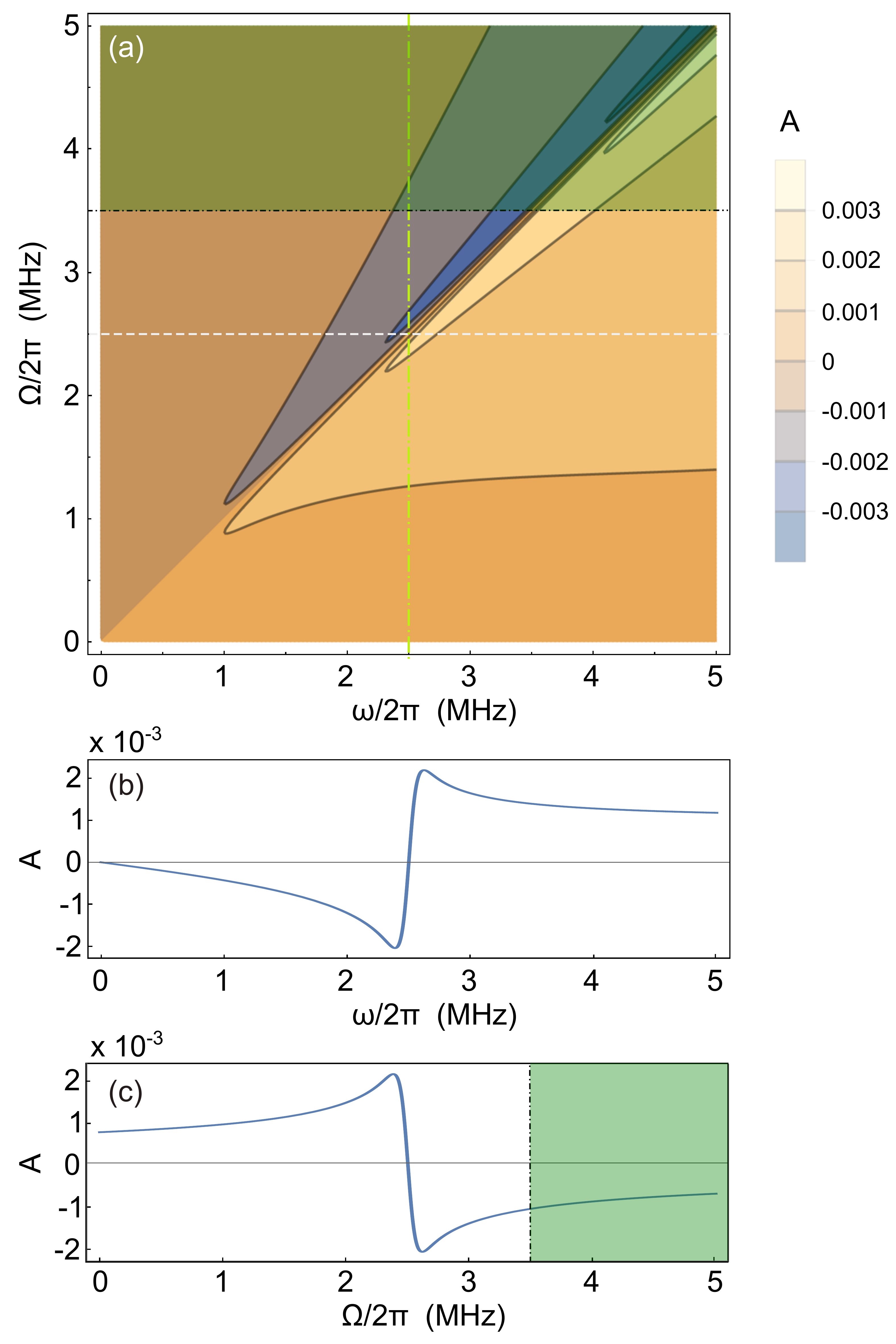}
\caption{(a) Rotational dissipation $A$ as function of the magnetic field and sphere frequencies $\omega/2\pi$ and $\Omega/2\pi$. (b) $A$ vs $\omega/2\pi$ for a fixed sphere rotation frequency $\Omega/2\pi=2.5$ MHz of the particle (white dashed line in Fig.~\ref{fig:2}a). (c) $A$ vs $\Omega/2\pi$, fixing the field frequency $\omega/2\pi=2.5$ MHz of the particle (green dot-dashed line in Fig.~\ref{fig:2}a). In panel (a) and (c) the dot-dashed black line denotes the micro-sphere maximum rotations frequency $\Omega_{max}/2\pi=3.5~$MHz and the green shaded area denotes the inaccessible region of frequencies. The parameters used are: $\sigma=5 \times 10^{9}$~Ohm$^{-1}$ m$^{-1}$, $\mu_r=1$, $N=40$, $Z_0=50$~Ohm, $R=50$~$\mu$m, $\ell=74.8$ m, $\epsilon_r=6.34$.}
\label{fig:2}
\end{figure}

The total energy stored in the EM mode is $E=1/2\left(  L_0 \ell +4L\right) I_0^2$, 
where the first term is the energy stored in the transmission line and the second term is the energy stored in the 4 inductors.  $L_0 =\sqrt{\varepsilon_r Z_0^2/c^2}$ is the transmission line inductance per unit length, where $\varepsilon_r$ is the transmission line permittivity. 
The dissipation $A$, which is the key quantity we are interested in, can be finally expressed as the inverse of a quality factor, $Q$:
\begin{equation}  
A = Q^{-1}= \frac{W}{\omega E} = \frac{ 8 \beta^2 \chi''}{(L_0 \ell+ 4L )}.
\label{G}
\end{equation} 
Under frequency inversion due to the Zel'dovich condition, $\chi''$ and hence also $A$ changes sign, so positive frequencies imply $A>0$ while negative frequencies imply $A<0$, i.e. gain.\\ 
We consider a rotating conductive micro-sphere with radius $R=50$ $\mu$m and conductivity $\sigma\sim 5 \times 10^{9}$ Ohm$^{-1}$m$^{-1}$, i.e. similar to that of common copper (residual resistance ratio $\approx 100$) at $T=4-10$ K \cite{website}. These parameters lead to rotational frequencies in the MHz range. This choice derives from a trade-off between conflicting requirements in the proposed experiment: a not too small sphere size to be compatible with existing fabrication and levitation technologies, and a high enough frequency $\omega$ to use transmission lines with realistic length.
Figure~\ref{fig:2} shows the predicted dissipation $A$ induced by the micro-sphere obtained as a function of the EM frequency, $\omega$, and rotation frequency, $\Omega$. 
Figure~\ref{fig:2} clearly shows a transition from absorption to amplification when $\omega$ approaches the sphere rotation frequency $\Omega$, thus showing evidence of a relatively strong Zel'dovich amplification that should be readily observable. 
Indeed, for $\omega>\Omega$, $A>0$, i.e. the particle absorbs the magnetic radiation impinging on it as expected. On the other hand, for $\omega<\Omega$, $A$ becomes negative, and hence the $\bm{B}$ field is amplified when scattered from the rotating sphere.\\
%
Fixing a feasible micro-sphere rotation frequency at $\Omega/2\pi=2.5$ MHz, we predict a maximum negative dissipation (i.e. gain) of $A_{max}\simeq -2\times10^{-3}$ (see Fig.~\ref{fig:2}b showing a line-out from Fig.~\ref{fig:2}a along the white dashed line at $\Omega/2\pi=2.5$ MHz).
We note that the chosen frequency is below the fundamental limit to the maximum rotation frequency $\Omega_{\mathrm{max}}$ that is determined by the sphere breaking apart. The breaking frequency scales with the inverse of radius and for $R=50$ $\mu$m we estimate $\Omega_{\mathrm{max}}/2\pi\approx 3.5$ MHz, indicated in Fig.~\ref{fig:2}a by the dot-dashed black line. The red shaded area denotes the physically inaccessible region of frequencies due to the breaking of the micro-sphere \cite{schuck2018ultrafast}. The scaling of the dissipation $A$ as function of $R$ can be found in the SM \cite{supply}.\\
Figure \ref{fig:2}c shows the dissipation $A$ trend for a fixed  circuit frequency $\omega/2\pi=2.5$ MHz as function of the micro-sphere rotation frequency $\Omega/2\pi$. \\
{\bf{Proposed experiment design details.}} In an experiment, the EM mode frequency of $2.5$ MHz can be obtained with a transmission line (see Fig.~\ref{fig:1}) formed by a compact coplanar waveguide of total length $\ell= |2\pi c/ (\omega\sqrt{\varepsilon_r})|\sim 74.8$ m, where $c$ is the speed of light and $\varepsilon_r=6.38$ for a silicon substrate. Each section of the circuit will need to be $\ell/4=18.7$ m long. 2-m long coplanar waveguides are routinely fabricated in superconducting Kinetic Inductance Traveling Amplifiers (KIT), on a typical chip area of $20 \times 20$ mm$^2$ \cite{Ranzani2018, Eom2012}. These circuits are typically based on a double spiral geometry with constant pitch $p$, for which the area $S$ is approximately proportional to the spiral length $d$ by the relation $S=pd$. By scaling existing spiral design \cite{Eom2012} a length of $18$~m corresponds to a maximum radius of $3.8$~cm. This is well within the current industrial standard for silicon wafers. This size is also compatible with standard commercial cryostats \cite{website4}.\\ 
In order to evaluate the experimental observability of the amplification,  we need to also compare $A_{max}$ with all the other sources of dissipation, $A_0$, in the electromagnetic modes, such as circuit losses. 
Meander-like superconductive coplanar waveguides (i.e. the KITs proposed above) or micro-coaxial niobium cables can be assumed to have $A_0^{wg}\sim 10^{-5}$~\cite{o2008microwave,zmuidzinas2012superconducting,kurpiers2017characterizing}. Furthermore, micro-coils in LC circuits show an intrinsic dissipation upper limit of $A_0^{coils}\sim 10^{-4}$ \cite{Gottardi2019}. These contributions are all more than one order of magnitude smaller than the gain predicted in Fig.~\ref{fig:2} and are not expected to therefore contribute appreciably. \\ 
Other materials for the sphere may also be considered. For example, in the Supplementary Material \cite{supply} we show that a smaller amplification, of the order of $10^{-5}$, can be achieved by magnetic particles of ferrite with moderate permeability $\mu_r\approx 900$ \cite{website3}. A similar analysis for the case of simultaneously conductive and magnetic materials (data not shown) leads to lower gain factors in comparison to the purely conductive case presented here. 
In the dielectric case, reported in the Supplementary Material \cite{supply}, $A$ is 5 orders of magnitude smaller due to limitations in the electric field amplitude that can be generated by capacitors (which now substitute the magnetic field and inductors, respectively).\\
We further observe that the particle does not need to be fully metallic. It can be a composite with a ferromagnetic core which allows an easier levitation \cite{maglev} and a highly conductive coating such that the coating thickness is comparable to the penetration depth of the EM field (see Ref.~\cite{supply}). \\
%
{\bf{Conclusions.}} 
We propose a superconducting circuit combined with a free-space, rotating sphere that is coupled to the circuit: this provides efficient EM-sphere coupling at low EM frequencies which in turn allow to access experimentally feasible rotational frequencies for the sphere, as required by the Zel'dovich condition. \\
Our calculations show that the proposed set-up exhibits measurable amplification of EM waves from mechanical rotation of both metallic and magnetic particles. It is worth noting that if GHz rotation frequencies could be reached with a similar scheme, then the circuit can be cooled to milli-kelvin temperatures where the thermal population is negligible, i.e. $k_b T < \hbar \omega$, a condition which has enabled superconducting quantum electrodynamics (QED) \cite{QED} and the study of fundamental physical effects such as dynamical Casimir emission \cite{casimir,casimir2}.  
GHz rotation frequencies have been obtained with optical trapping of sub-micron sized spheres \cite{GHZ,tong}. This would require to arrange 4 nano-coils at distance $\sim 100$ nm from the particle, which is incompatible with a realistic trapping laser waist. Moreover, the combination of a high power laser beam with a milli-Kelvin environment would require careful control of heat dissipation aspects. However, if these technological issues can be solved, spontaneous Zel'dovich emission could then also be observed i.e. rotational generation of photons out of the quantum vacuum \cite{zel1971generation} in a superconducting QED experiment. In addition, the system proposed here relies on a new generation of superconducting circuits coupled to rapidly moving elements, thus allowing the study of further fundamental physics problems such as the detection of rotational quantum friction \cite{Pendry2012,Abajo2010,Kardar2012} and quantum vacuum friction \cite{Davies2005}.\\
\begin{acknowledgments}
{\bf{Acknowledgments.}} DF and MCB acknowledge financial support from EPSRC (UK Grant No. EP/P006078/2) and the European Union's Horizon 2020 research and innovation programme under Grant Agreement No. 820392. HU, AV and GG acknowledge financial support from the EU H2020 FET project TEQ (Grant No. 766900) and the Leverhulme Trust (RPG-2016-046). AV thanks Iacopo Carusotto for helpful discussions. 
\end{acknowledgments} 



\onecolumngrid
\newpage
{\centering{\large \bfseries Zel'dovich amplification in superconducting circuits: supplementary information\par}\vspace{2ex}
	{ Maria Chiara Braidotti$^{1,*}$, Andrea Vinante$^{2,3}$, Giulio Gasbarri$^{2}$, Daniele Faccio$^{1}$, Hendrik Ulbricht$^{2,*}$\par}
{\centering  \small \emph{$^{1}$School of Physics and Astronomy, University of Glasgow, G12 8QQ, Glasgow, UK.\\ 
$^{2}$Department of Physics and Astronomy, University of Southampton, SO17 1BJ, Southampton, UK.\\
$^{3}$Istituto di Fotonica e Nanotecnologie - CNR and Fondazione Bruno Kessler, I-38123 Povo, Trento, Italy.}\par}}

\par\vspace{.5ex}

\renewcommand{\theequation}{S\arabic{equation}}
\renewcommand{\thefigure}{S\arabic{figure}}
 \setcounter{equation}{0}
\section{Conductive micro-sphere}
\subsection{Levitated micro-sphere gas damping.}
We analyse in detail the damping, i.e. slowing down of a levitated spinning micro-sphere due to friction originating from gas in the surrounding environment.\\
When the electromagnetic driving system (that puts the micro-sphere in rotation) is switched off, the micro-sphere rotation frequency $\Omega$ decreases over time due to the scattering with gas molecules, 
according to the law \cite{gabis1996, epstein1924}
\begin{equation}
{d\Omega\over dt} = -\frac{\Omega}{\tau} \qquad\mbox{with}\qquad \tau=\frac{\pi}{10}\frac{\rho v a}{P},
\label{rate}
\end{equation}
where $\tau$ is the rate of attenuation. In eq. (\ref{rate}) $\rho$ is the micro-sphere density, $P$ is the gas pressure and $v=\sqrt{\frac{8 k_b T}{\pi m}}$ is the average speed of the gas molecules with mass $m$ at temperature $T$, $k_b$ is the Boltzmann constant. \\
Figure \ref{fig:1} shows the attenuation rate $\tau$ trend with pressure $P$ for a Copper micro-sphere ($\rho=8960$ Kg/m$^3$) of radius $a=50$ $\mu$m, as  considered in the manuscript, levitated in a helium gas (mass $m=6.68\times10^{-27}$ Kg) at temperature $T\simeq 4.2$ K. We observe that for pressures of the order of $10^{-5}$ mbar, the rate of decrease of the rotation frequency is $\bar{\tau}\simeq 2\times 10^4$ s. We conclude that the these conditions will provide a time window of order of 1 hour at a time to perform measurements. For example, in 36 minutes the micro-sphere slows down by only 10\%, thus providing a sufficiently  constant rotation rate. \\ 
\begin{figure}[h]
\includegraphics[scale=.4]{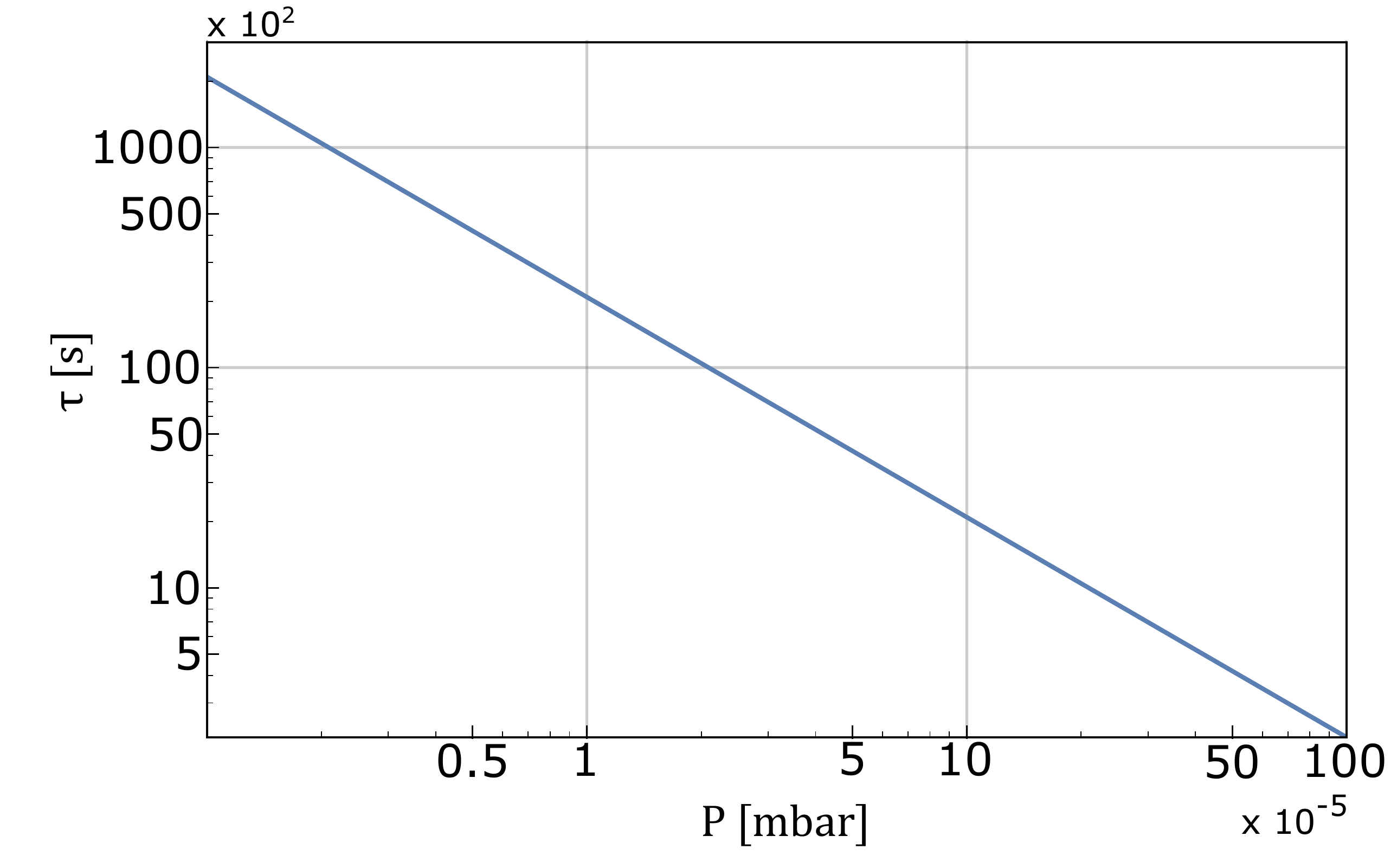}
\caption{Rate of attenuation $\tau$ as a function of the gas pressure $P$. At $P=10^{-5}$ mbar, the attenuation rate is $\bar{\tau} = 2\times10^4$ s, meaning that the micro-sphere frequency $\Omega$ is attenuated of a factor $1/e$ after more than 5.5 hours.}
\label{fig:1}
\end{figure}
\subsection{Dipole moment $\bm{m_0}$ and dissipation $A$}
Here, we compute the magnetic dipole moment $\bm{m_0}$ induced by the magnetic field on the Conductive micro-sphere from which we can then compute the dissipation $A$. As in the manuscript and following \cite{reichert2012complete,schuck2018ultrafast}, we consider the spinning sphere axis oriented along $z$ and the magnetic flux density $\bm{B_0}$ produced by the coils to be uniform (see Eq. (3) of the manuscript).\\
All the following calculations are carried out in the reference frame co-rotating with the micro-sphere. In this frame, the field generated by the coils is $\bm{B_r}= 2\beta \bm{b_r}I_j$, where $I_j$ is the current in the j-th lumped inductor, $ \bm{b_r} = (1,i,0)^T e^{i\omega_{r}t}$ with $\omega_r=\omega-\Omega$, where $\omega$ and $\Omega$ are the frequency of the field $\bm{B_0}$ and the rotation frequency of the micro-sphere in the lab frame, respectively. Note that the $x$ and $y$ components of $\bm{b_r}$ are phase shifted by $\pi/2$ in order to provide a magnetic field with angular momentum (OAM $q=1$). \\
Our purpose is to find the form of the vector potential $\bm{\mathcal{A}}$, which allows us to compute the magnetic dipole moment and the dissipation $A$. 
To this aim, we solve Maxwell's equations, which describe the interaction of the magnetic field $\bm{B_r}$ generated by the circuit and the conductive micro-sphere. Solutions to the Maxwell's equations with appropriate boundary conditions for this system can be derived analytically as shown in \cite{reichert2012complete}. \\
We choose the Coulomb gauge ($\nabla\cdot\bm{\mathcal{A}}=0$) where we have
\begin{eqnarray}
\nabla\times \bm{E} &=& -\frac{\partial\bm{B}}{\partial t} \label{ME1_1} \\
\nabla\times \bm{B} &=& \mu_0\mu_r\bm{J} \label{ME1_2}
\end{eqnarray}
where $\bm{E}$ and $\bm{B}$ are the electric and magnetic fields in the whole space respectively, while $\mu_0$ and $\mu_r$ are the vacuum and relative permeability. $\bm{J}$ is the current density given by the conduction current and the displacement current. We consider only the conductor current, with $\bm{J}=\sigma\bm{E}$, where $\sigma$ is the micro-sphere conductivity as the displacement contribution to the current is negligible for metals at the frequencies we are considering in the manuscript. 
From Eq. (\ref{ME1_1}) and the vector potential definition $\nabla\times\bm{\mathcal{A}}=\bm{B}$ we find 
\beq
\bm{E}=-\frac{\partial\bm{\mathcal{A}}}{\partial t}.
\eeq
Manipulating equations (\ref{ME1_1}) and (\ref{ME1_2}) and using the relation $\nabla\times(\nabla\times\bm{\mathcal{A}})=\nabla(\nabla\cdot\bm{\mathcal{A}})-\Delta\bm{\mathcal{A}}$, we find the equation for the vector potential
\begin{equation}
-\Delta\bm{\mathcal{A}}+\mu_0\mu_r\sigma\frac{\partial\bm{\mathcal{A}}}{\partial t}=0.
\label{eqA}
\end{equation}
In order to solve this last equation for $\bm{\mathcal{A}}$, we divide the space in two regions: one inside the micro-sphere with radius $R$ and one outside it. At the interface of the two areas, we consider the following three boundary conditions:
\begin{eqnarray}
B_{\perp}^{in} &=& B_{\perp}^{out} \label{bc1}\\
H_{\parallel}^{in} &=& H_{\parallel}^{out} \label{bc2}\\
\bm{\mathcal{A}} \underset{r\rightarrow\infty}{\longrightarrow} &\bm{\mathcal{A}_{\infty}}&=\frac{1}{2}\bm{B_r}\times\bm{r} \label{bc3}
\end{eqnarray}
where $\bm{B}=\mu_0\mu_r\bm{H}$ and $r$ is the position vector in the co-rotating reference frame centred at the center of the micro-sphere. The magnetic field outside the micro-sphere $\bm{B^{out}}=\bm{B_r}+\bm{B^{refl}}$ is composed of two components: the incident magnetic field generated by the circuit $\bm{B_r}$ plus the magnetic field reflected from the micro-sphere $\bm{B^{refl}}$. The magnetic field inside the micro-sphere $\bm{B^{in}}$ is given by the component of the incident field transmitted in the scattering.\\
Considering the last boundary condition, the complex vector potential $\bm{\mathcal{A}}$ can be generally written as 
\begin{equation}
\bm{\mathcal{A}} = \frac{1}{2}\left[F(r)\bm{B_r}\times\bm{r}\right]
\label{A}
\end{equation}
where the function $F(r)$, such that $F(r) \to \infty =1$, is as yet unknown. By replacing the vector potential ansatz (\ref{A}) in eq. (\ref{eqA}), we find two equations for $F(r)$, one for the space inside the sphere of radius $R$ and one for the surrounding space, which are:
\begin{eqnarray}\label{eq_f}
&&\frac{\partial^2 F(r)}{\partial r^2} + \frac{4}{r}\frac{\partial F(r)}{\partial r} - i\mu_0\mu_r\sigma\omega_r F(r)=0 \qquad\mbox{for}\quad R\leq r\\
&&\frac{\partial^2 F(r)}{\partial r^2} + \frac{4}{r}\frac{\partial F(r)}{\partial r} =0 \qquad\mbox{for}\quad R>r
\end{eqnarray}
with solutions, 
\begin{eqnarray}
&&F(r) = c_1\left(\frac{\sin(c_{EC}r)}{(c_{EC}r)^3}-\frac{\cos(c_{EC}r)}{(c_{EC}r)^2}\right) = c_1 f(c_{EC}r) \qquad\mbox{for}\quad R\leq r\\
&&F(r) = c_2+\frac{c_3}{r^3}\qquad\mbox{for}\quad R>r
\end{eqnarray}
where $c_{EC}=\sqrt{-i\mu_0\mu_r\sigma\omega_r}$. In order to determine the unknown constants $c_1$, $c_2$ and $c_3$, we observe that the boundary conditions for $\bm{B}$ and $\bm{H}$ translate in conditions for the function $F(r)$:
\begin{eqnarray}
&&F^{out}(r\rightarrow R) = F^{in}(r\rightarrow R)\\
&&\mu_r\left[r\frac{\partial F^{out}(r)}{\partial r} + 2F^{out}(r)\right]_{r\rightarrow R}=\left[r\frac{\partial F^{in}(r)}{\partial r} + 2F^{in}(r)\right]_{r\rightarrow R}\\
&&F(r) \rightarrow 1 \qquad\mbox{for}\quad r\gg R
\end{eqnarray}
which lead to
\begin{eqnarray}
&&F(r) = 1+D(c_{EC}R)\left(\frac{R}{r}\right)^3 \qquad\mbox{for}\quad R\leq r\\
&&F(r) = \left[1+D(c_{EC}R)\right]\frac{f(c_{EC}r)}{f(c_{EC}R)}  \qquad\mbox{for}\quad R>r.
\end{eqnarray}
The function $D$ is
\begin{eqnarray}
&&D(c_{EC}R) = \frac{(2\mu_r+1)g(c_{EC}R)-1}{(\mu_r-1)g(c_{EC}R)+1}    \\
&&g(c_{EC}R) = f(c_{EC}R)\frac{c_{EC}R}{\sin(c_{EC}R)}=\frac{1-(c_{EC}R)\cot(c_{EC}R)}{(c_{EC}R)^2}.
\end{eqnarray}
These last equations define the explicit form of the vector potential $\bm{\mathcal{A}}$ in the whole space. \\
It is now possible to compute the form of the magnetic dipole moment $\bm{m_r}$ on the micro-sphere in the reference frame co-rotating with the micro-sphere. 
From the relation between $\bm{m_r}$ and the the reflected component of the vector potential $\bm{\mathcal{A}}$
\beq\label{AvsM}
\bm{\mathcal{A}}=\frac{\mu_0}{4\pi}\frac{\bm{m_r}\times\bm{r}}{r^3}
\eeq
we find
\beq\label{magn_dipole}
\bm{m_r}=\frac{2\pi}{\mu_0} I_j R^3 \beta [F(R)-1]\begin{pmatrix}
1 \\
i \\
0
\end{pmatrix} e^{i\omega_r t}= \left(\chi'+i\chi''\right) \bm{B_r}.
\eeq
where $\chi=\chi'+i\chi''$ is the complex response function of the sphere in the presence of the field in the co-rotating reference frame. It is worth noting that in order to compute the amount of reflected flux into the coil, we subtracted a factor $1$ from $F(R)$  corresponding to the contribution of the incident field $\bm{B_r}$. \\ 
The torque exerted on the particle is given by the cross product $\bm{T}=\Re\{\bm{m_0}\}\times\Re\{\bm{B_0}\}=-2 I_j \beta \Re\{m_{\perp}\}\hat{z}$, where $m_{\perp}$ is the component of $\bm{m_0}$ perpendicular to $\bm{B_0}$ which gives the back-reflection on the circuit: 
\begin{eqnarray}\label{m_yZ}
m_{\perp} &=& \beta I_j \frac{2 R^3 \pi }{\mu_0} \left[i(F(R)-1)\right]e^{i\omega t} = \\
		&=& \beta I_j \frac{2 R^3 \pi }{\mu_0} \left[i\frac{(2\mu_r+1) \left(1-Rc_{EC}\cot\left[Rc_{EC}\right]\right)-R^2 c_{EC}^2}{(\mu_r-1) \left(1-Rc_{EC}\cot\left[Rc_{EC}\right]\right)-R^2 c_{EC}^2}\right]e^{i\omega t}.
\end{eqnarray}
From this last equation we can compute the field dissipation (Eq. (6) in the manuscript) caused by the presence of the micro-sphere in the system.
The dissipation $A$ is given by
\beq
A = \frac{4\alpha''}{L_0l + 4L} = \frac{4}{L_0l + 4L}\frac{2 \pi \beta^2R^3}{\mu_0} \Re\left[i\frac{(2\mu_r+1) \left(1-Rc_{EC}\cot\left[Rc_{EC}\right]\right)-R^2 c_{EC}^2}{(\mu_r-1) \left(1-Rc_{EC}\cot\left[Rc_{EC}\right]\right)-R^2 c_{EC}^2}\right]
\eeq
where we used the relation $\alpha''= 2 \beta^2\chi''$, derived in the manuscript.

\subsubsection{Amplification as effective negative resistance}
An equivalent picture of the amplification process is obtained by looking at the process from the point of view of the coils. Here we set for simplicity $I_j=I$. The rotating magnetic moment will couple a magnetic flux back into each coil with peak value $\Phi = \beta |\bm{m_0}| $. We can thus write a relation $\Phi = \alpha I$, with $\alpha = \alpha'+i\alpha''$: the flux response function is given by $\alpha = 2 \beta^2 \chi$. The imaginary part $\alpha''(\omega_r)= 2 \beta^2 \chi''(\omega_r)$ describes a flux in quadrature with the current $I$, or equivalently a voltage $V=-i\omega \Phi$ in phase with $I$, corresponding to a resistance with mean dissipation $W_c=\Re\{VI\}/2=\omega \alpha'' I_0^2/2 = \omega \beta^2 \chi'' I_0^2$, where the factor $1/2$ comes from averaging over one cycle. Summing over the 4 coils leads to Eq.~($5$) in the manuscript. The factor $\omega \alpha''$ can be seen as a resistance $\mathcal{R}$ induced by the rotating sphere into the coil, hence amplification corresponds to a negative resistance that leads to power emission into the EM mode as opposed to the expected (for a non-rotating or slowly rotating sphere) power absorbed from the EM mode. 


\subsection{Penetration depth}
When a metal is immersed in a magnetic field, the electromagnetic (EM) radiation can penetrate inside the material. From the Beer-Lambert law, the intensity of the EM radiation inside the material falls as $I(r)=I_0e^{-r/\delta_p}$, where $r$ is the the coordinate orthogonal to the metal surface  of the micro-sphere and $I_0$ is the intensity of the field outside it. The  ``penetration depth" $\delta_p$ defines the radiation attenuation length at which radiation is damped by a factor $1/e$ and depends on the properties of the material as $\delta = \sqrt{2/(\sigma\mu_0\mu_r \omega)}$ \cite{FeymanBook}.\\
We can conclude that there is no need to have a micro-sphere that is fully conductive since the field will only penetrate up to $\delta_p\simeq 11$ $ \mu$m in the considered system with radius $R=50 \mu$m, conductivity $\sigma=5\times 10^9$ Ohm$^{-1}$ m$^{-1}$, $\mu_r=1$ and $\omega/2\pi=2.5$ MHz. Moreover, Refs. \cite{reichert2012complete,schuck2018ultrafast} show that the penetration depth is further reduced with increasing spinning frequency $\omega_r$, i.e. the frequency of the field with respect to the micro-sphere.\\
Figure \ref{fig:new2}a reports the analysis of the dissipation $A$ obtained for different frequencies $\omega_r$ in the  co-rotating reference frame and micro-sphere radii $R$. Let us remark that $A<0$ correspond to an amplification of the electromagnetic wave.
The maximum gain (negative dissipation) arises when the penetration depth $\delta_p$ is of the same order of the sphere size $R$, being that $R\propto \omega_r^{-1/2}$ as can be seen from the definition of the penetrations depth $\delta_p$.
From this, it is possible to find the optimal conductivity for our system, which allows to chose the best material to perform the experiment. Figure \ref{fig:new2}b shows $A$ as function of the conductivity $\sigma$ for a microsphere of radius $R=50~\mu$m as that chosen in the manuscript. It is worth to notice that there is a maximum negative dissipation at approximately $\sigma=5 \times 10^9$ Ohm$^{-1}$m$^{-1}$, showing that Copper is indeed an optimal candidate. 

\begin{figure}[h]
\includegraphics[scale=.25]{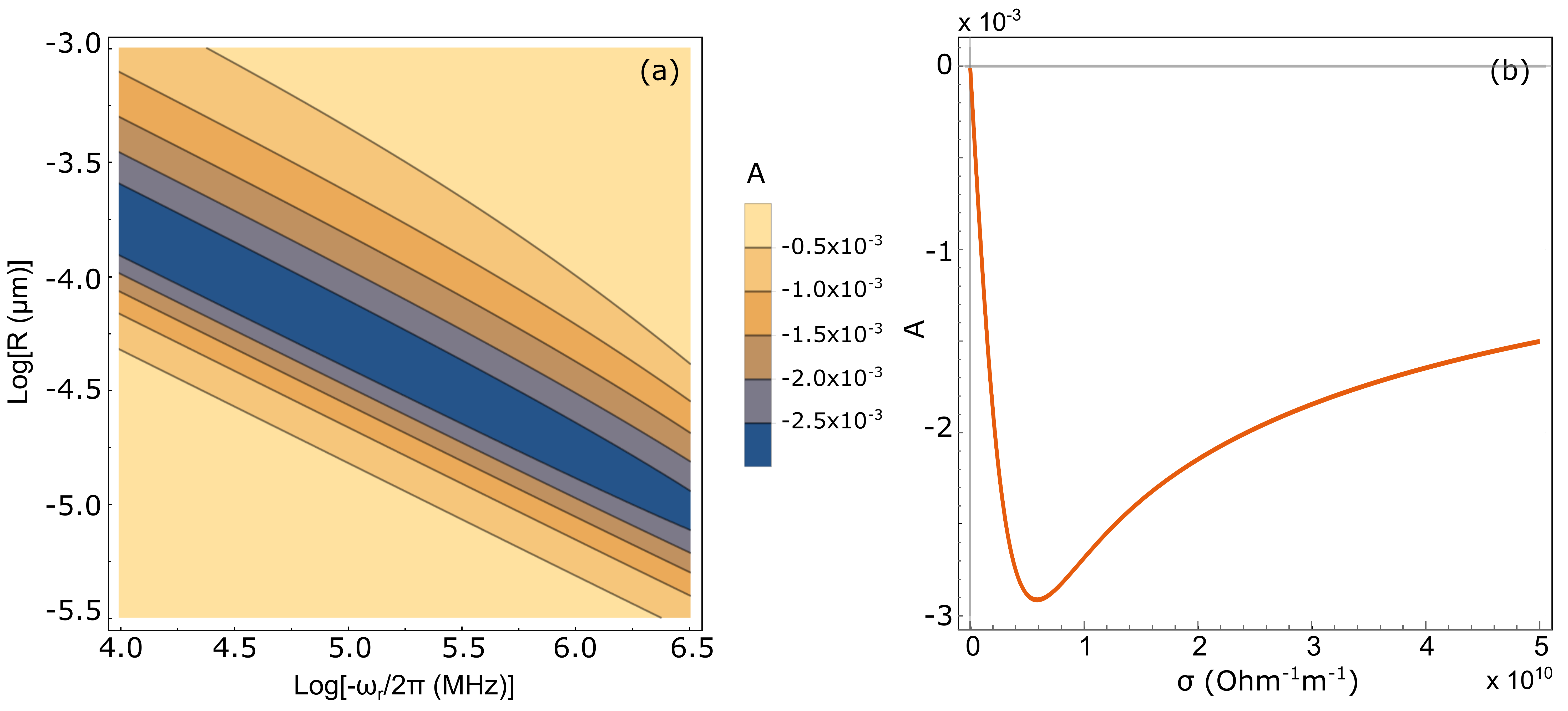}
\caption{(a) Rotational dissipation $A$ as function of the magnetic field frequency in the corotating reference frame $\omega_r/2\pi = (\omega-\Omega)/2\pi$ and micro-sphere radius $R$ in log-log scale. The parameters used are: $\sigma=5 \times 10^{9}$~Ohm$^{-1}$ m$^{-1}$, $\mu_r=1$, $N=40$, $Z_0=50$~Ohm, $\ell=74.8$ m, $\epsilon_r=6.34$. One can see that the optimal radius scales as $\omega_r^{-1/2}$, which is the expected dependence of the penetration depth. (b) Dissipation $A$ as function of the conductivity $\sigma$ for a micro-sphere of radius $R=50~\mu$m for a fixed frequency $\omega_r/2\pi=-10^5$ Hz. Same parameters as in panel (a).}
\label{fig:new2}
\end{figure}

%

\subsection{Effect of coil thickness}
The amount of Zel’dovich amplification in the proposed system depends on the geometry of the inductors in the circuit, being that the dissipation $A\propto\beta^2$, where the parameter $\beta$ depends on the geometry of the coils and their distance from the micro-sphere (see Eq.~(3) in the manuscript). A quantitative estimate of the parameter $\beta$ can be calculated from the Biot-Savart formula $\beta = \frac{N I \mu_0 R^2}{2 \sqrt{R^2+h^2}}$, where $R$ is the coil radius and $h$ is the distance of the coil from the centre of the micro-sphere. Here we have neglected the coil wire thickness and have approximated the coil as $N$ loops of radius $R$ (see Fig. S3). In the proposed circuit geometry we approximated $h\approx R$ (see also Fig. 1 of the manuscript), to maximize the Zel’dovich effect. However, the thickness of the coil wire can be considered replacing the parameter $\beta$ by an effective $\beta_{\text{eff}}$ averaged over different loops. By calculating $\beta_{\text{eff}}$ it is possible to estimate how the coil thickness affects the Zel’dovich amplification. 
\begin{figure}[h]
\includegraphics[scale=.35]{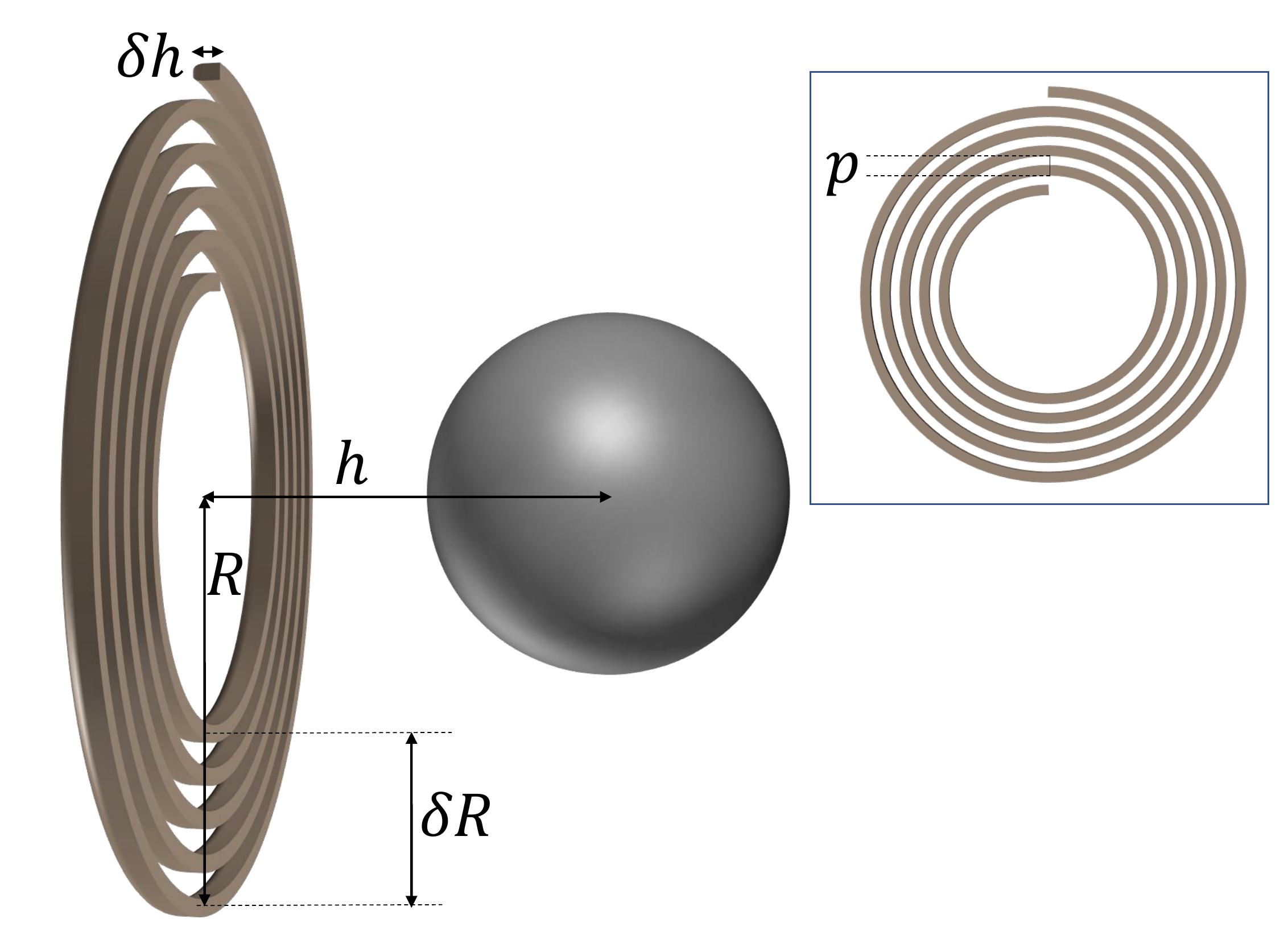}
\caption{3D Planar inductor sketch. $h$ defines the distance between the center of the microsphere and the coil. $R$ is the radius of the microsphere. $\delta h$ and $\delta R$ are the coil thickness along $h$ and along the radius respectively. The inset shows the planar coil pitch $p$. }
\label{fig:new2}
\end{figure}
Two types of thickness have to be considered: (i) along $h$ ($\delta h$) and (ii) along $R$ ($\delta R$), both due to the section of the wire. If the section of the coil is much smaller than $h$ $(\delta h\ll h)$, and the radial thickness is much smaller than $R$ $(\delta R\ll R)$, then $\beta\simeq\beta_{\text{eff}}$ and the correction to the dissipation due to the thickness of the coil are negligible. Such conditions is satisfied by a planar coil with external radius $R=100~\mu$m, $N=40$ and wire thickness $\delta h=100~$nm and pitch $p=300~$nm. Planar coils with these characteristics can be fabricated by electron-beam lithography techniques \cite{coils}. The coil thicknesses are : (i) $\delta h=100~$nm and (ii) $\delta R= p N=12~\mu$m. From these parameters it is possible to compute the $\beta_{\text{eff}}$ parameter being $\beta_{\text{eff}} = \frac{N I \mu_0 (R-\delta R/2)^2}{2 \sqrt{(R-\delta R/2)^2+(h+\delta h/2)^2}}\approx 0.97 \beta$, which reduces the gain $A$ by $5\%$.

\section{Ferro-Magnetic micro-sphere ($\sigma=0$)}
\subsection{Dipole moment $\bm{m_0}$ and dissipation $A$}
\begin{figure}[h]
\includegraphics[scale=.4]{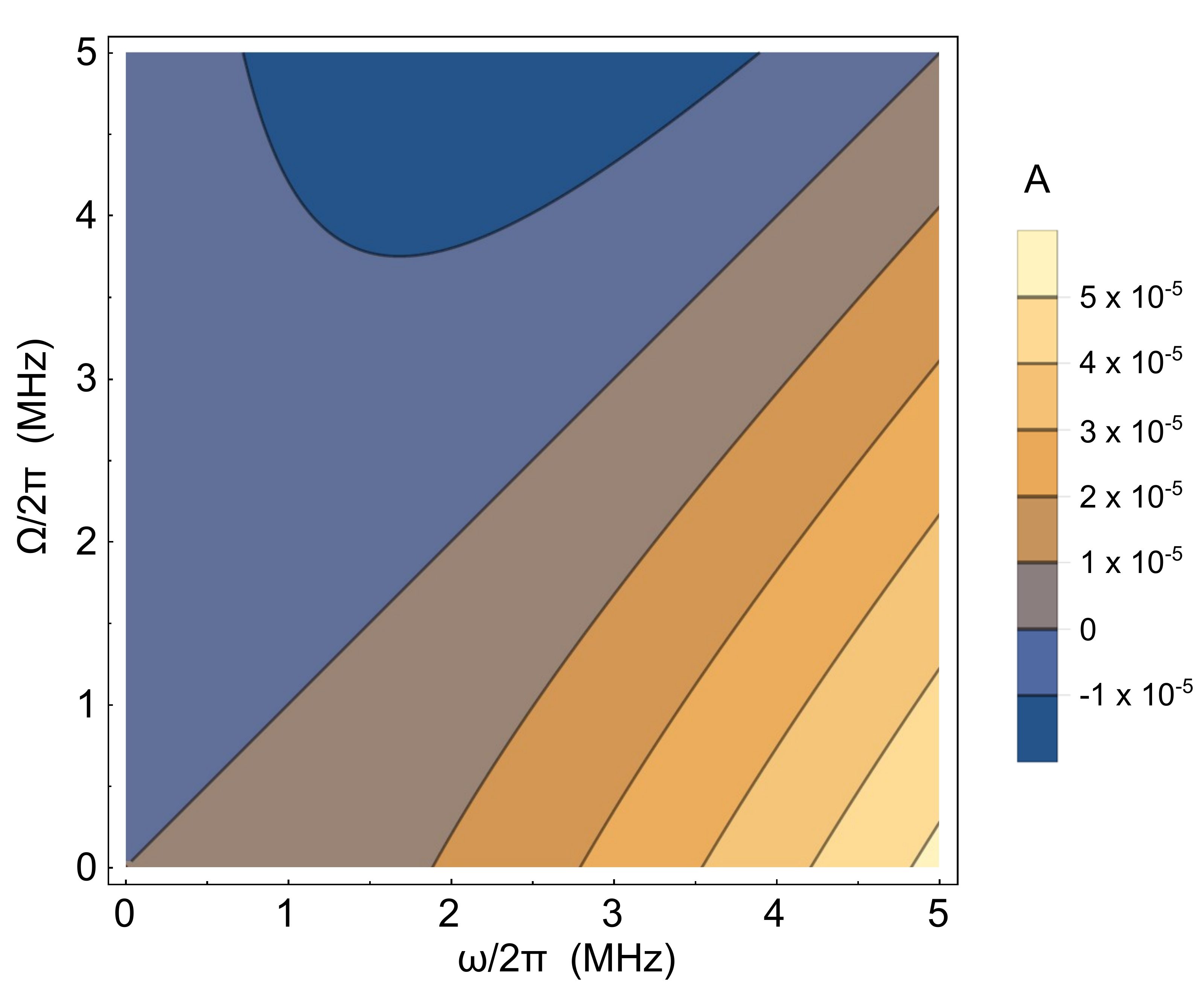}
\caption{Rotational dissipation $A$ as a function of the magnetic field oscillation and micro-sphere rotation frequencies, $\omega$ and $\Omega$. The parameters used are:
$\sigma=0$ Ohm$^{-1}$m$^{-1}$, $\mu_i=900$, $\omega_{\mu}=6$ MHz, $N = 40$, $Z_0 = 50$ Ohm, $R = 50$ $\mu$m, $\ell = 74.8$ m, $\epsilon_r = 6.34$.}
\label{fig:3}
\end{figure}

We consider a ferro-magnetic micro-sphere. The calculations follow those derived in the previous section. The only difference is in the form of the $F(r)$ function. With $\sigma=0$ for a ferro-magnetic and non-conductive material, Eqs. (\ref{eq_f}) become the same inside and outside the micro-particle, i.e.,
\begin{eqnarray}
&&\frac{\partial^2 F(r)}{\partial r^2} + \frac{4}{r}\frac{\partial F(r)}{\partial r} =0 \qquad\mbox{for}\quad R\leq r\\
&&\frac{\partial^2 F(r)}{\partial r^2} + \frac{4}{r}\frac{\partial F(r)}{\partial r} =0 \qquad\mbox{for}\quad R>r.
\end{eqnarray}
Their solution is
\begin{eqnarray}
&&F(r) = d_1+\frac{d_2}{r^3} \qquad\mbox{for}\quad R\leq r\\
&&F(r) = d_3+\frac{d_4}{r^3} \qquad\mbox{for}\quad R>r.
\end{eqnarray}
From the boundary conditions (\ref{bc1}-\ref{bc3}) for $\bm{B}$ and $\bm{H}$ we find
\beq
d_1= 1+\frac{2(\mu_r-1)}{2+\mu_r}, \qquad d_2=0, \qquad d_3=1, \qquad d_4=\frac{2(\mu_r-1)}{2+\mu_r}.
\eeq
From this we can write 
\begin{eqnarray}
&&F^{in}(r) = 1+\frac{2\mu_r-2}{2+\mu_r} \left( \frac{R}{r} \right)^3 \qquad\mbox{for}\quad R\leq r\\
&&F^{out}(r) = \frac{3\mu_r}{2+\mu_r}  \qquad\mbox{for}\quad R>r.
\end{eqnarray}
With the same  the relation Eq. (\ref{AvsM}) between $\bm{m_r}$ and the vector potential $\bm{\mathcal{A}}$, 
we find  
\begin{eqnarray}\label{m_y}
\bm{m_r} = \frac{4 \pi }{\mu_0} \left[\frac{(\mu_r(\omega_r)-1)}{2+\mu_r(\omega_r)}R^3 \right]\bm{B_r}=\chi\bm{B_r},
\end{eqnarray}
whose perpendicular component $m_{\perp}\propto\sin(\theta(\omega_r))$ where $\theta\propto\arg\left[\left( \frac{(\mu_r(\omega_r)-1)}{2+\mu_r(\omega_r)}\right)R^3 \beta\right]$. In this case the torque is $ T\propto-|\bm{B_0}|\sin(\theta)$, which changes sign for negative $\omega_r$, allowing Zel'dovich amplification, due to the property $\mu_r(-\omega_r)=\mu_r^*(\omega_r)$ \cite{LandauBook_stat}. The permeability $\mu_r(\omega_r)$ can be often written as a one-pole response function
\beq
\mu_r(\omega)=\frac{\mu_i}{1+i\omega_r/\omega_{\mu}},
\eeq
where $\mu_i$ and $\omega_{\mu}$ are the initial permeability and the maximum usable frequency, respectively. 
The dissipation $A$ results to be
\beq
A = \frac{4\alpha''}{L_0l + 4L} = \frac{4}{L_0l + 4L} \beta^2 \frac{4 \pi }{\mu_0} \Re\left[\frac{i(\mu_r-1)}{2+\mu_r}R^3\right].
\eeq
Figure \ref{fig:3} shows the dissipation $A$ as function of EM field frequency  and of the micro-sphere rotation frequency in the magnetic sphere case. The material chosen is a commercial  MnZn ferrite \cite{website2} with $\mu_i=900$ and $\omega_{\mu}=6$ MHz at room temperature ($T=300$ K). The amplification factor is of the order of $10^{-5}$ which is $2$ orders of magnitude smaller than that obtained with a metallic micro-sphere, which was $10^{-3}$. It is worth noticing that the properties of ferrites change significantly with temperature and  at cryogenic temperatures $\mu_i$ decreases significantly, i.e. the micro-sphere becomes more absorptive \cite{cosier1967} with an increase of the dissipation $A$. The effect could be increased if magnetic materials with lower initial permeability and/or lower cut-off frequency $\omega_{\mu}$ were available.

\section{Dielectric case.}
\begin{figure}[h]
\includegraphics[scale=1.5]{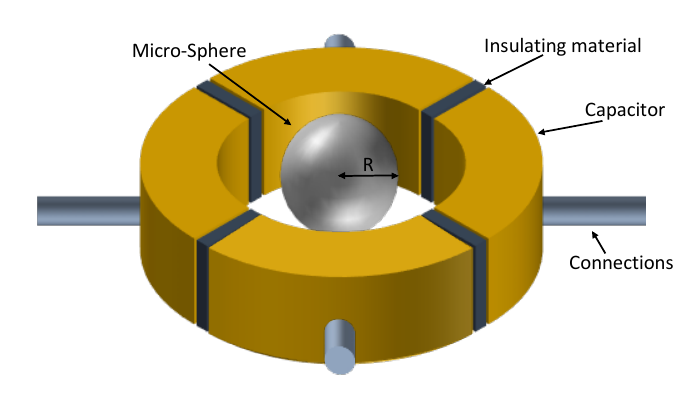}
\caption{Experimental set-up: the circuit is divided in 4 sections of length $\ell/4$, connected by 4 electrodes with capacity $C$ to form a closed ring configuration. The distance between each capacitor and the center of the sphere is $2R$.}
\label{fig:4}
\end{figure}

In the case of a dielectric particle an electric probe field plays the role of the magnetic field analysed in the previous cases. Figure \ref{fig:4} shows the experimental setup proposed to generate a rotating electric field $E_0$. Inductors are substituted by 4 electrodes forming a ring around the spinning micro-sphere. The following calculation is carried out in the reference frame co-rotating with the micro-sphere. In analogy with the previous cases, in the reference frame co-rotating with the micro-sphere the electric field inside the cavity can be written as $\bm{E_r} = \frac{V}{2 R} \bm{b_r}$, where $V$ is the electric potential applied to each electrode. Each electrode face is $2 R$ apart from the center of the micro-sphere.
The calculation is analogue to that of the magnetic case however the vector potential is replaced by the scalar electric potential $\phi$. The starting equation is the Laplace equation 
\beq\label{laplace}
\nabla^2\phi=0.
\eeq
whose solutions in spherical coordinates are
\beq
\phi=\sum_{n=1}^{\infty} \left(	A_n r^n+\frac{\beta_n}{r^{n+1}}\right)P_n(\cos(\theta)).
\eeq
with $r$ the radial coordinate and $\theta$ the azimuthal coordinate. $A_n$ and $\beta_n$ are coefficients which depend on the boundary conditions and $P_n$ are the Legendre polynomials. \\
As in the previous case we have to account for the boundary conditions in order to find the values of the coefficients $A_n$ and $\beta_n$. We solve the Laplace Eq. (\ref{laplace}) by considering the following boundary conditions:
\begin{eqnarray}
&&\phi^{out}(r\rightarrow\infty)=|\bm{E_r}| r \cos(\theta) \\
&&\phi^{in}(r=0) \mbox{ has not to diverge} \\
&&\phi^{out}(a)=\phi^{in}(R) \label{be3}\\
&& \epsilon_r\partial_r\phi^{in}(R)=\partial_r\phi^{out}(R).\label{be4}
\end{eqnarray}
From the first and second equations, we find $A^{out}_1=-|\bm{E_r}|$, $A^{out}_{n\neq1}=0$ and $\beta^{in}_{n}=0$, which substituted in Eq. (\ref{be3}) gives
\beq
-|\bm{E_0}| r\cos(\theta)+\sum_{n=1}^{\infty} \frac{\beta^{out}_n}{r^{n+1}}P_n(\cos(\theta))=\sum_{n=1}^{\infty}	A^{in}_n r^n P_n(\cos(\theta)).
\eeq
Comparing this last equation order by order, we obtain 
\begin{eqnarray}
&&\beta_1^{out}= (A_1^{in}+|\bm{E_r}|)R^3\\
&&A^{out}_{n\neq1} R^{2n+1} =\beta^{out}_{n\neq1}.
\end{eqnarray}
Substituting this last equation in Eq. (\ref{be4}) we obtain
\beq
\epsilon_r \sum_{n=1}^{\infty} \left(	A^{in}_n n R^{n-1}\right)P_n(\cos(\theta))=\sum_{n=1}^{\infty} \left(	A^{out}_n n R^{n-1}+(n+1)\frac{-\beta^{out}_n}{R^{n+2}}\right)P_n(\cos(\theta))
\eeq
where we took into account that $\beta^{in}_{n}=0$. Comparing order by order we obtain
\begin{eqnarray}
&& A_1^{in} = -\frac{3|\bm{E_r}|}{\epsilon_r+2}\\
&&\beta_{n\neq1}^{out}= 0\\
&&\beta^{out}_{1} = \frac{\epsilon_r-1}{\epsilon_r+2}|\bm{E_r}| R^3.
\end{eqnarray}
From this we can write the scalar field $\phi$ inside and outside the sphere as 
\begin{eqnarray}
&&\phi^{in} = \frac{-3|\bm{E_r}|}{\epsilon_r+2}r\cos(\theta)\\
&&\phi^{out}= -|\bm{E_r}| r\cos(\theta)+|\bm{E_r}|\left(\frac{\epsilon_r-1}{\epsilon_r+2}\right)\frac{R^3}{r^2}\cos(\theta).
\end{eqnarray}
The interaction between the electric probe field and the rotating sphere induces a complex electric dipole moment $\bm{d}$ on the sphere given by 
\beq \label{d_y}
\bm{d} = 4\pi\epsilon_0 \left[ \left(\frac{\epsilon_r-1}{\epsilon_r+2}\right)  R^3\right] \bm{E_r}=\chi\bm{E_r},
\eeq
where $\epsilon_0 = 8.854 \times 10^{-12}$ F/m is the permittivity of vacuum and whose component perpendicular to the electric field is proportional to the $d_{\perp}\propto\sin(\theta)$ with $\theta\propto\arg\left[\left( \frac{\epsilon_r-1}{2+\epsilon_r}\right) R^3E\right]$. As in the previous case the torque on the sphere is $ T\propto-|\bm{E_0}|\sin(\theta)$, which changes sign for negative $\omega_r$, allowing Zel'dovich amplification, due to the property $\epsilon_r(-\omega_r)=\epsilon_r^*(\omega_r)$ \cite{LandauBook_stat}. 
\begin{figure}[t]
\includegraphics[scale=.4]{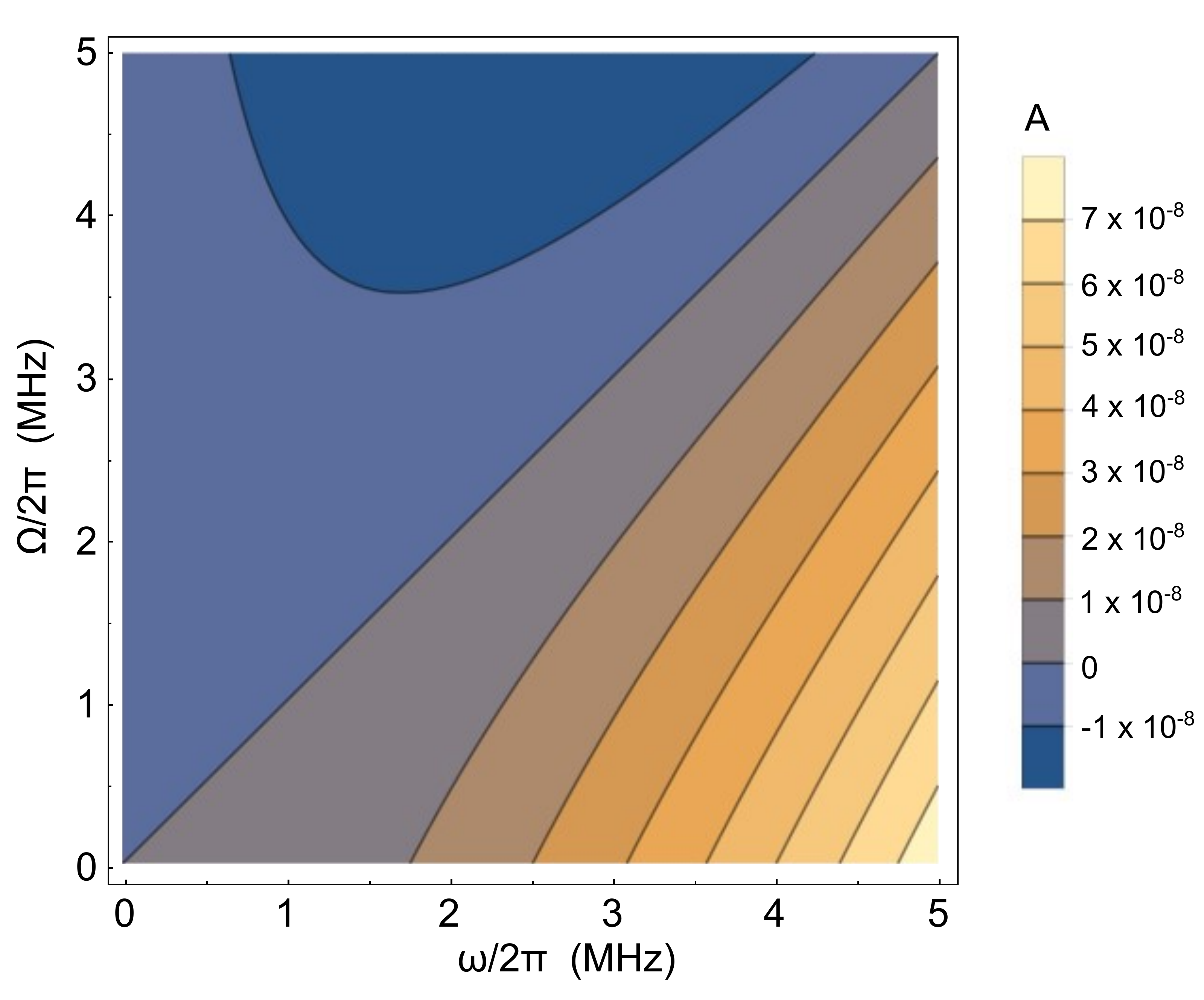}
\caption{Dielectric rotational dissipation $A$ as function of the electric field oscillation and micro-sphere rotation frequencies, $\omega$ and $\Omega$. The parameters used are: $R=50$ $\mu$m, $\epsilon_{\infty}=4.6$, $\epsilon_i= 73.9$ and $\omega_{\epsilon}=0.1$ MHz, $Z_0 = 50$ Ohm, $C = \epsilon_0 2 R$, $\ell = 74.8$ m, $\epsilon_r = 6.34$.}
\label{fig:5}
\end{figure}
The dielectric constant varies with frequency as \cite{midi}
\beq \label{DC}
\epsilon_r(\omega_r)=\epsilon_{\infty} + \frac{ \epsilon_i}{1+2 i\omega_r/\omega_{\epsilon}}.
\eeq
where $\epsilon_{\infty}$ is the high-frequency dielectric constant while $\epsilon_i$ and $\omega_{\epsilon}$ are the initial permittivity and the loss peak frequency respectively which vary among different materials.\\
The charge reflected from the micro-sphere is hence given by $Q_b = -d_{\perp} E_0/V = (\alpha'+i\alpha'') V $ which gives a dissipation 
\beq
A = \frac{4\alpha''}{C_0\ell + 4C} = \frac{4}{C_0\ell + 4C} \frac{E_0^2}{V} 4\pi\epsilon_0 \left(\frac{\epsilon_r-1}{\epsilon_r+2}\right)R^3.
\eeq
where $C_0$ is the capacity of the transmission line. 
Figure \ref{fig:5} shows the dielectric dissipation $A$ as function of the oscillation frequency of the field $\omega$ and of the micro-sphere rotation $\Omega$ in the dielectric case. In order to maximize the effect we choose to levitate a particle of water which has a high absorption at $T=22$ $^{\circ}$C in the MHz frequency range with  $\epsilon_{\infty}=4.6$, $\epsilon_i= 73.9$ and $\omega_{\epsilon}=0.1$ MHz \cite{midi}. 
The resulting amplification factor is of the order of $10^{-8}$, several orders of magnitudes below the magnetic and metallic cases. As explained in the manuscript this huge difference is due to the fact that inductors are more efficient in the creation of magnetic fields compared to capacitors as a result of the fact that the intensity of the magnetic field increases with the number of loops in inductors.


\end{document}